\documentclass{aa}
\usepackage{times}
\usepackage{amssymb}
\usepackage{graphicx}
\usepackage{natbib}

\begin{document}

\thesaurus{06(%
08.09.2 \mbox{Cyg X-1}; 
08.02.1; 
13.25.5)} 

\title{Temporal Evolution of X-ray Lags in Cygnus~X-1}  

\author{K.~Pottschmidt\inst{1} \and J.~Wilms\inst{1} \and
  M.A.~Nowak\inst{2} \and W.A.~Heindl\inst{3} \and D.M.~Smith\inst{4} \and
  R.~Staubert\inst{1}} 
\institute{Institut f\"ur Astronomie und Astrophysik -- Astronomie,
  University of T\"ubingen, Waldh\"auser Str. 64, D-72076 T\"ubingen, 
  Germany
\and
  JILA, University of Colorado, Boulder, CO 80309-440, U.S.A.
\and
  Center for Astronomy and Space Sciences, Code 0424, University of
  California at San Diego, La Jolla, CA 92093, U.S.A.
\and
  Space Sciences Laboratory, University of California at Berkeley,
  Berkeley, CA 94720, U.S.A.} 

\offprints{K. Pottschmidt}
\mail{K. Pottschmidt (katja@astro.uni-tuebingen.de)}
\titlerunning{Lags in Cygnus X-1}
\authorrunning{K.~Pottschmidt et al.}
\date{Received $<$date$>$ / Accepted $<$date$>$ } 
\maketitle

\begin{abstract}
  We present the long term evolution of the fre\-quency-depen\-dent X-ray
  time lags of the black hole candidate \mbox{\object{Cygnus X-1}} as
  measured in 1996 and 1998 with the Rossi X-ray Timing Explorer (RXTE).
  Lag spectra measured during the 1996 June soft state are very similar to
  those seen during 1996 December and most of 1998 while \mbox{Cyg~X-1} was
  in its hard state. During state transitions, however, the shape and
  magnitude of the X-ray lag is highly variable and tends to be much larger
  than outside of the state transitions. This behavior is most obvious in
  the 1--10\,Hz band. The increase of the X-ray lag during the state
  transitions might be related to the formation and destruction of the
  synchrotron radiation emitting outflows present during the hard state.
\end{abstract}

\section{Introduction}\label{sec:intro}

Galactic black hole candidates (BHC) are predominantly found in two generic
states: the hard state, in which the X-ray spectrum is a Comptonization
spectrum emerging from a hot electron cloud with a typical electron
temperature of $\sim$150\,keV \citep{dove:97b,poutanen:98a}, and the soft
state, in which the X-ray spectrum is thermal with a characteristic
temperature of $kT_{\rm BB}\lesssim 1$\,keV to which a steep power-law is
added \citep[and references therein]{cui:96a,gierlinski:99a}. Transitions
between these states have been seen in all persistent galactic BHC, with
the exception of \mbox{\object{LMC~X-1}}. Optically thick radio emission is
observed during the hard state, during transitions between the hard and the
soft states the radio emission tends to be optically thin and more highly
variable. Finally, during the soft state, galactic black hole candidates
tend to be radio quiet \citep[see, e.g.,][for a
review]{fender:00b}. Although the phenomenology of the states is rather
well understood, the accretion geometry in these sources is still a matter
of debate. Most current models for the hard state posit an accretion disk
corona with a large covering factor that Comptonizes most of the accretion
disk radiation \citep[and references therein]{poutanen:98a}. Its physical
size is also assumed to be rather large. On the other hand, the corona is
assumed to have almost vanished during the soft state, where the X-ray
luminosity is dominated by thermal radiation \citep{gierlinski:99a}.

In recent years, several attempts have been made to use X-ray timing
methods to constrain these models. In addition to the power spectrum
analysis \citep{belloni:90a,gilfanov:99b}, higher order statistics like the
frequency-dependent coherence function and time lags have proven to be
useful in evaluating physical accretion models \citep{hua:98a,nowak:98a} by
providing combined spectral and temporal information. For example, the
maximum expected time delay between hard and soft photons is roughly the
size of the Comptonizing medium divided by the slowest propagation speed of
a disturbance, while the minimum time lag is roughly the photon diffusion
time through the corona \citep{nowak:98c}.

The canonical BHC, \mbox{Cygnus~X-1}, stays predominantly in the hard
state, but occasionally transits into the soft state for a few months
\citep[][see also
Fig.~\ref{fig:softlag}]{gierlinski:99a,cui:96a,cui:97c}. The apparent
decrease of the X-ray lags during the 1996 state transitions was considered
evidence that the size of the accretion region during the soft state is
smaller than during the hard state \citep{cui:97c}. First comparisons of
transition and soft state lags to hard state lags, however, indicated that
the physical interpretation has to be more complex \citep{cui:99a}.

In addition to the state transitions, quasi-regular variations on a
$\sim$150\,d timescale are observed in the X-ray spectral shape and soft
X-ray flux as well as in other wavelength bands
\citep{pooley:98a,brocksopp:99b}. In 1998 we initiated a monitoring
campaign with the Rossi X-ray Timing Explorer (RXTE) to systematically
study the multi-wavelength long term variation of \mbox{Cyg~X-1} over a
period of several years. During 1998, weekly pointings of $\sim$3\,ks
duration were performed. In later years, longer exposure times (but larger
sampling intervals) were used \citep{pottschmidt:00a}. In this Letter we
present results of the first year of the campaign, focusing on the X-ray
time lags. We describe our data analysis methods (Section~\ref{sec:data})
and compare the temporal behavior of \mbox{Cyg~X-1} during 1998 with that
seen during 1996 (Section~\ref{sec:tempevol}). Specifically, we show that
large X-ray lags appear to be associated with \emph{transitions} between
the soft and hard state, but \emph{not} with the state itself. In
Section~\ref{sec:disc} we discuss implications for the accretion models in
galactic black hole candidates.

\begin{figure}

\resizebox{\hsize}{!}{\includegraphics{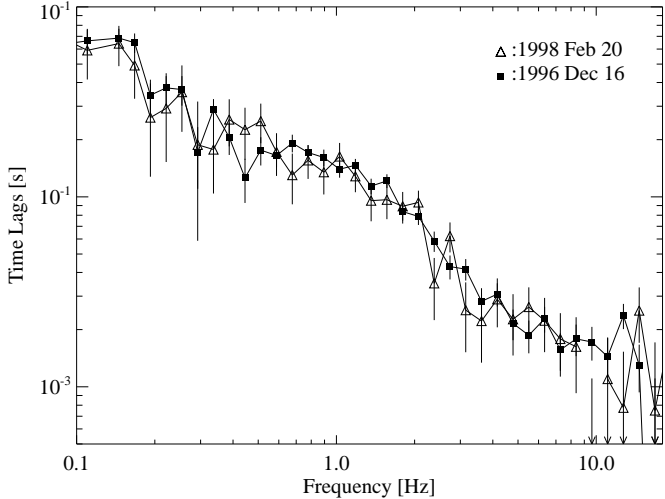}}

\caption{Comparison of the X-ray lag spectrum for one of the 3\,ks
monitoring observations of 1998 with that obtained for $\sim$30\,ks of the
RXTE observation of 1996 December 16. The lags have been measured between
$\lesssim$4\,keV and $\sim$8--13\,keV. Both examples are typical hard state
observations.}
\label{fig:cmpoldnew}
\end{figure}

\section{Observations and Data Analysis}\label{sec:data}
The RXTE data presented here were obtained with the Proportional Counter
Array \citep[PCA;][]{jahoda:96b} and with the All Sky Monitor
\citep[ASM;][]{remillard:97a}. We used the standard RXTE data analysis
software, ftools~4.2.  A log of the observations is presented in Table~1
which is available in electronic form only from the Centre de Don\'ees
Stellaires (CDS).  The data were reduced using the procedures described in
detail in our analysis of the RXTE observations of
\mbox{\object{GX~339$-$4}} \citep{wilms:98c}.  Intervals with large
background flux were removed after visually inspecting the ``electron
ratio''.  As a result of the data screening, $\sim$2\,ks of usable data
were left for each of the 3\,ks monitoring observations.  We then extracted
lightcurves with a resolution of 16\,ms for three energy bands
($\lesssim$4\,keV, $\sim${}8--13\,keV, and 18.3--72.9\,keV)\footnote{Since
the PCA data modes used to obtain the 1996 data differ from those used in
1998, it was impossible to use identical energy bands for all
observations. The bands used to compare the 1996 data to the 1998 data were
the closest matches possible (the highest energy band is unavailable for
some of the 1996 soft state data). The detailed bands are given in Tab.~1
available from the CDS.}. The computation of the time lags for these energy
bands follows \citet{nowak:96a} and \citet{nowak:98a}.

According to our previous experience, the methods used to compute the
uncertainty of the time lag spectrum are applicable over a large range of
source fluxes and exposure times. Since most of the 1998 observations were
quite short, however, we independently verified the determination of the
lag in these cases by comparing them to much longer observations. As an
example, Fig.~\ref{fig:cmpoldnew} displays the time lag spectra for two RXTE
observations spaced by 1.25\,years. Taking the much larger uncertainty
from the short observation into account, the overall agreement is
excellent. To further increase the signal to noise ratio in the lag
determination we rebinned the X-ray lag spectrum into five frequency bands.

\section{Temporal Evolution of the Lag}\label{sec:tempevol}

\begin{figure}
\resizebox{\hsize}{!}{\includegraphics{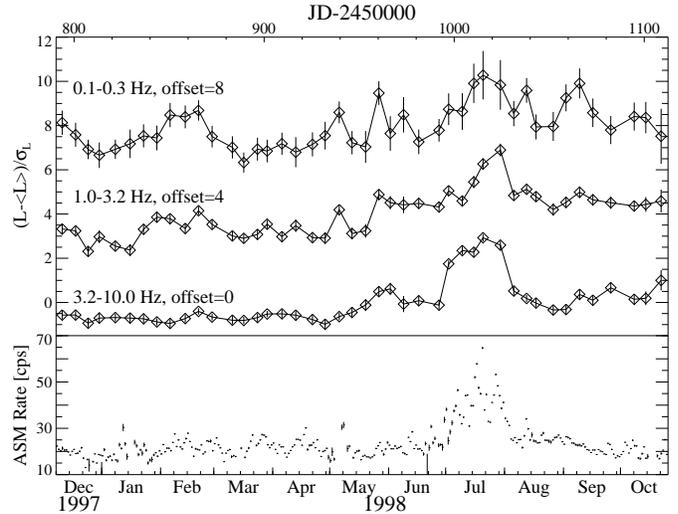}}

\caption{Diamonds: Temporal evolution of the average lag between
$\lesssim$4\,keV and 18.3--72.9\,keV for the indicated frequency
intervals. The deviation of the lag from its mean value as determined from
all observations of 1998 for the respective frequency band, in units of its
standard deviation is shown. Dashes: RXTE ASM 2--10\,keV count rate binned
to a resolution of 1\,d.}
\label{fig:lagtempevol}
\end{figure}

In Fig.~\ref{fig:lagtempevol} we display the evolution of the average time
lag for three representative frequency bands and the RXTE ASM count rate.
The Figure shows that a clear long term variability of the lags is present
during 1998. For frequencies above $\sim$1\,Hz, the mean lag is
significantly \emph{larger} during the interval in 1998~July that is
characterized by a larger ASM count rate. At lower frequencies the
fractional change of the lag decreases (Fig.~\ref{fig:lagtempevol}). This
tendency is a consequence of the temporal evolution of the shape of the lag
spectrum as characterized by its rough $f^{-\alpha}$ proportionality:
During 1998~July, $\alpha${}$\sim$0.6, compared to its usual value of
$\alpha${}$\gtrsim$0.7 \citep{nowak:98a}.

\begin{figure*}
\resizebox{12cm}{!}{\includegraphics{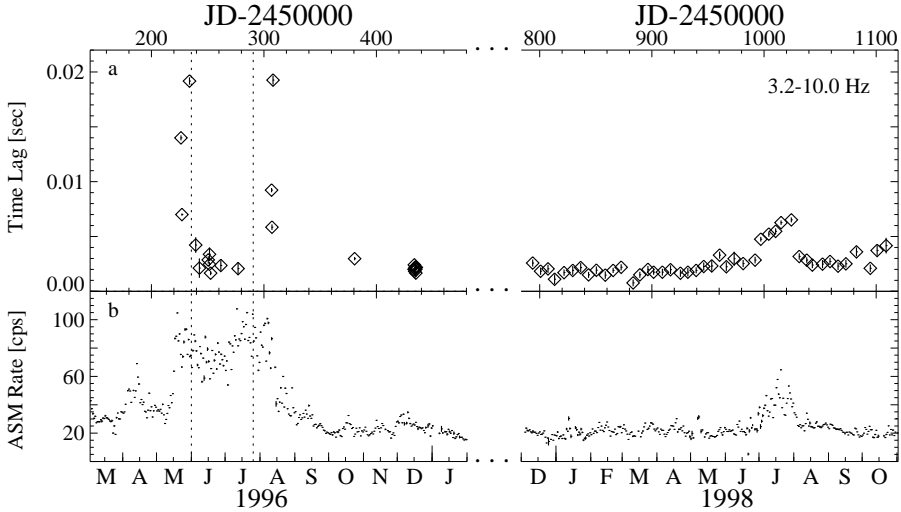}}
\hfill
\parbox[b]{55mm}{%
\caption{\textbf{a} Temporal evolution of the absolute value of the average lag
  between $\lesssim$4\,keV and $\sim$8--13\,keV in the 3.2 to 10\,Hz band
  and \textbf{b} ASM count rate for 1996 and 1998. The dotted lines denote
  the 1996 soft state.\label{fig:softlag}} }
\end{figure*}

\begin{figure*}
\resizebox{12cm}{!}{\includegraphics{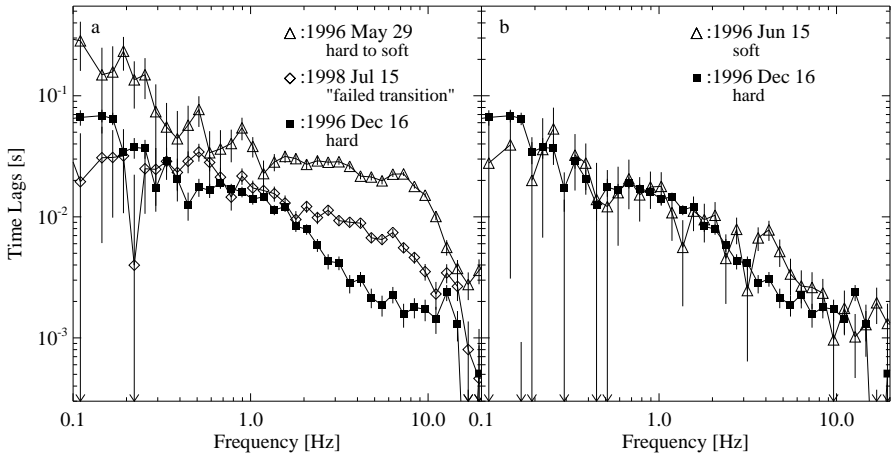}}
\hfill
\parbox[b]{55mm}{%
\caption{X-ray lag spectra between $\lesssim$4\,keV and
$\sim$8--13\,keV. \textbf{a} Comparison of the hard to soft state
transition, the 1998 July transitional state, and the typical hard
state. \textbf{b} Comparison of the 1996 soft state and the hard state.} 
\label{fig:lagshape}} 
\end{figure*}

Compared to the hard state, intervals like 1998~July show an increased disk
contribution to the X-ray spectrum \citep{zdz:99a,gilfanov:99b} and have
often been associated with ``failed state transitions''. One would expect
the lag behavior to show the same tendency as during a successful state
transition. At first glance, the 1996 data presented by \citet{cui:97c}
suggest that we should expect the 1998~July lags to \emph{decrease} during
this interval. This contradicts our results (Fig.~\ref{fig:lagtempevol}).
We therefore went back to the 1996 soft state data and applied the same
analysis as for the 1998 data. We also computed X-ray lags for two
observations performed after the 1996 soft state, one in 1996 October
\citep{dove:97c,nowak:98a}, as well as the one in 1996 December
\citep{focke:98a}. Fig.~\ref{fig:softlag} shows the absolute values of the
average lag between $\lesssim$4\,keV and $\sim$8--13\,keV for 1996 and
1998. The lags are indeed longer during the state transitions than they are
in the soft state, however, during the soft state itself, the absolute
value of the X-ray lag equals that of the hard state. In fact, the
frequency dependence of the lag is very
similar for the soft and hard state (Fig.~\ref{fig:lagshape}b).
Previous analyses of two hard state observations already suggested that the
soft and hard state lags might not be so different as previously thought
\citep{cui:99a}. Our numerous hard state monitoring observations now
clearly indicate that \emph{the X-ray lag spectrum of \mbox{Cyg~X-1} is
rather independent of the spectral state}. The \emph{enhanced lags are then
associated with transition or failed transition intervals}, and not with
the state of the source itself.

\section{Discussion and Conclusions}\label{sec:disc}
In Fig.~\ref{fig:lagshape}a we display examples for the whole range of lag
spectra present in this analysis.  Taking the shape of the typical hard
state lag spectrum as a baseline, the lag is significantly longer during
the 1998~July failed state transition. The frequency range from 1 to 10\,Hz
exhibits these changes most prominently. During the hard to soft transition
in 1996~May, the lag is longer by almost a factor 10 at 6\,Hz. As already
shown in section~\ref{sec:tempevol}, these results indicate that the
magnitude and shape of the X-ray lag spectrum in \mbox{Cyg~X-1} is related
to the state transitions.

Previous models for the generation of the X-ray lags assumed more or less
static media to produce the X-ray lag by scattering of seed photons in a
Comptonizing medium. The size of the region required to produce the
observed lags in such a model is large \citep[$\gtrsim$300 gravitational
radii;][]{nowak:98c}. Such a large size simplifies models in which the soft
state Compton corona is much smaller than the hard state corona, as had
been initially inferred from the 1996 soft state lags and spectral
shape. Indirect evidence for the change of the size of the X-ray emitting
corona has also been presented by \citet{zdz:99a}, \citet{gilfanov:99b},
and commented on by \citet{dimatteo:99a}, with the latter authors
suggesting an upper limit for the hard state coronal radius of $\sim$30
gravitational radii. \citet{zdz:98a} suggested that in the hard state of
GX~339$-$4 there is a correlation between the X-ray power law photon index
and the fraction of this power law that is reflected by cold
material. Specifically, they suggested that softer power laws implied
greater reflection, which implies smaller coronae in certain models. Within
the hard state, softer power laws are associated with higher luminosities,
implying again that the corona is shrinking as the source goes from the
hard to the soft state. Our data clearly show that the soft and hard state lag
spectra are very similar. This makes the geometrical interpretation of the
lags in terms of Comptonization models alone difficult. We therefore need
to look for other models to explain the observed lags.
In the following we present a
qualitative picture, based on recent observational results. Note,
however, that a detailed theoretical model is beyond the scope of this Letter.

In recent years, evidence has emerged that the state transitions are not
solely an X-ray phenomenon. Studies of \mbox{Cyg~X-1} and \mbox{GX~339$-$4}
have revealed that in the hard state the X-ray and radio behavior is
correlated \citep{brocksopp:99b,corbel:00a}. In \mbox{GX~339$-$4}, there is
strong evidence that the source is radio quiet during the soft state
\citep{fender:99b}. At least in one case, optically thin radio flares
accompanied the state transition. In \mbox{Cyg~X-1}, no radio data are
available for the soft state. The end of the 1996 soft state as well as the
1998 July ``failed state transition'', however, coincided with radio flares
\citep{zhang:97e,brocksopp:99b}. Such flaring events are typically
associated with the ejection of a synchrotron emitting cloud from the
central, X-ray emitting region \citep{corbel:00a,fender:99b}.  

We suggest that the scattering of primary X-rays in ejected material may be
responsible for the enhanced transition X-ray time lags in \mbox{Cyg~X-1}:
In the hard state, a stable, presumably partially collimated, radio
emitting outflow (``jet'') exists, while the soft state shows no outflow
\citep{brocksopp:99b}. During the formation and (failed) destruction of the
jet, when radio flaring is observed, outflows that are uncollimated and
much larger than those of the hard state might be present. Assuming that
the hard state lag spectrum is produced in the accretion disk \citep[which
still poses a problem for most models;][]{nowak:98c} and that the rather
weak hard state jet does not significantly affect these intrinsic lags, the
prolonged lags during the transitions could be produced in these large
ejected outflows. Additional scattering might also be responsible for the
reduced X-ray coherence that was reported by \citet{cui:97c} for the 1996
state transitions. Such a model wherein a fraction of the observed lags is
created near the base of a radio emitting jet or wind has been previously
suggested in analogy to blazar jet emission models \citep[van Paradijs
1999, priv.\ comm., see also ][]{fender:99b}. The state change following
the flaring/ejection could then lead to a new accretion disk configuration
with different X-ray spectral emission characteristics, different power
spectra, but similar inherent lag spectrum.


\begin{acknowledgements}
  This work has been financed by DFG grant Sta~173/22, by NASA grants
  NAG5-3072, NAG5-3225, and NAG5-7265, and by a travel grant from the
  DAAD. We are indebted to the RXTE schedulers, most notably E.~Smith, for
  making such a long monitoring campaign feasible.  We thank C.~Brocksopp,
  R.~Fender, P.~Kretschmar, I.~Kreykenbohm, and W.~Cui, the referee, for
  helpful comments.
\end{acknowledgements}


\end{document}